\documentclass[10pt,twocolumn,twoside]{IEEEtran}
\usepackage{amsmath,amssymb,amsfonts}
\usepackage{cite}
\usepackage{graphicx}
\usepackage{array}
\usepackage{algorithmicx}
\usepackage{algpseudocode}
\usepackage{setspace}
\usepackage{url}
\usepackage{epstopdf}
\usepackage{verbatim}
\usepackage{color}
\usepackage{bm}
\usepackage{multirow}
\usepackage[table,xcdraw]{xcolor}
\usepackage{booktabs}
\usepackage{makecell}
\usepackage{ntheorem}
\usepackage{textcomp}
\usepackage{mathrsfs}
\usepackage{subcaption}
\usepackage[utf8]{inputenc}

\newtheorem{Thm}{Theorem}

\newtheorem{Prob}{Problem}

\begin{document}

\title{Physics-Informed WiFi Sensing for Robust 3D Human Pose Estimation in Mobile and Cross-Environment Settings}

\author{
Kaixuan Huang, Yuanbo Chen, \IEEEmembership{Student Member, IEEE}, Guangjin Pan, \IEEEmembership{Member, IEEE}, \\ Shiyi Mu, Tao Yu, Guhan Zheng, Shunqing Zhang, \IEEEmembership{Senior Member, IEEE}
\thanks{
This work was supported by the Science and Technology Commission Foundation of Shanghai under Grant 24DP1500703, and the National Natural Science Foundation of China (NSFC) under Grant 62571307. (Corresponding Author: Shunqing Zhang)}
\thanks{Kaixuan Huang, Yuanbo Chen, Shiyi Mu, Guhan Zheng, Shunqing Zhang are with the School of Information and Communication Engineering, Shanghai University, Shanghai, 200444, China (e-mail: \{xuan1999, cyb24721331, mushiyi, guhanzheng, shunqing\}@shu.edu.cn).}
\thanks{Guangjin Pan is with Department of Electrical Engineering, Chalmers University of Technology, 41296 Gothenburg, Sweden (e-mail: guangjin.pan@chalmers.se)}
\thanks{Tao~Yu is with the Department of Electrical and Electronic Engineering, The University of Hong Kong, Hong Kong (e-mails:taoyee@hku.hk).}
}

\maketitle

\begin{abstract}

Device-free human pose estimation using commodity WiFi signals has emerged as a promising paradigm for pervasive sensing in mobile computing systems. However, existing approaches often suffer from severe performance degradation when deployed across heterogeneous environments, due to complex multipath propagation and domain shifts in wireless signals.
In this paper, we present a physics-informed WiFi sensing framework for robust 3D human pose estimation under mobile and cross-environment settings. Our approach explicitly models wireless signal propagation characteristics and incorporates multipath-aware attention to capture environment-dependent signal variations. To further improve generalization, we introduce a disentangled representation learning scheme that separates pose-related features from environment-specific factors, enabling effective cross-domain adaptation without requiring extensive retraining.
We implement our system using commodity WiFi devices and evaluate it on multiple public benchmarks, including Person-in-WiFi-3D and MM-Fi, as well as real-world deployments across diverse indoor environments. Experimental results demonstrate that our framework significantly improves robustness and generalization performance compared to state-of-the-art methods, particularly under cross-environment scenarios. These results highlight the potential of physics-informed wireless sensing for enabling reliable, scalable, and infrastructure-free human-centric applications in mobile computing systems.
\end{abstract}

\section{Introduction}

With the proliferation of commodity WiFi infrastructure, device-free human sensing has become an increasingly important capability in mobile computing systems. Unlike camera-based approaches~\cite{li2025adaptive, Li2021TokenPose, zhang2022mvpose}, wireless sensing enables privacy-preserving sensing without capturing visual information, making it suitable for deployment in sensitive indoor environments. In addition, it relies on existing wireless infrastructure and does not require users to carry dedicated devices. These advantages make wireless sensing attractive for a wide range of applications, including healthcare monitoring, smart environments, and human-computer interaction~\cite{wu2025enhancing, Zheng2023DeepLearningHPE}. 

\begin{figure}[t!]
\centering
\includegraphics[width= \columnwidth]{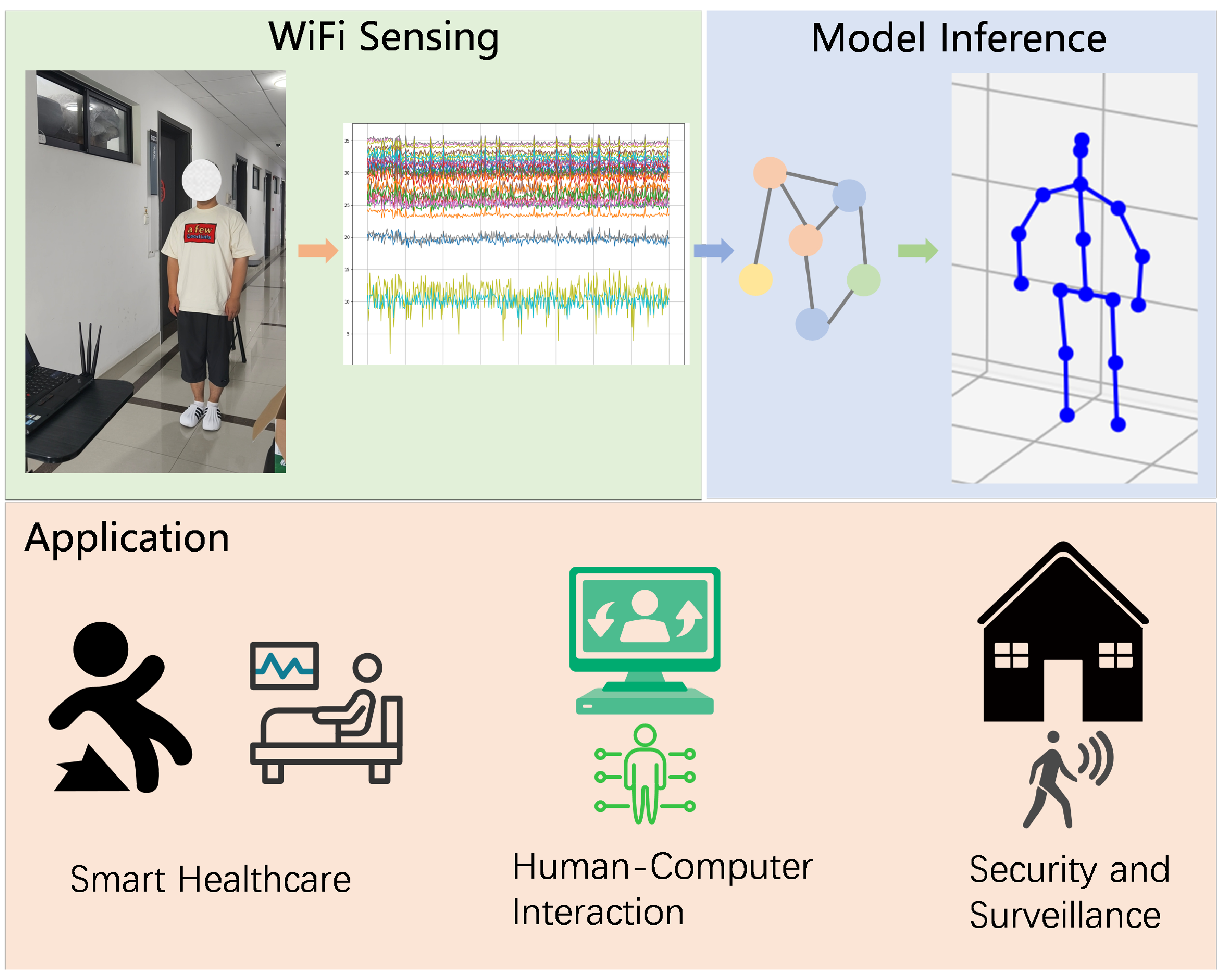}
\caption{WiFi-based pose estimation enables privacy-preserving sensing through walls and in darkness, while leveraging ubiquitous infrastructure for practical deployment in healthcare, smart homes, and elderly care.}
\label{fig:WiFi}
\end{figure}

Compared with specialized sensing modalities such as mmWave radar~\cite{deng2023midas++}, WiFi sensing can be deployed using widely available 802.11 equipment, which supports scalable and cost-effective deployment. In wireless systems, Channel State Information (CSI) characterizes the frequency response across multiple subcarriers and captures fine-grained multipath propagation induced by human motion. When signals propagate through the environment, they undergo reflection and scattering before reaching the receiver. This physical process enables pose estimation without line-of-sight and under challenging lighting conditions, making it particularly attractive for applications in hospitals, elderly-care facilities, and private homes. However, realizing this potential in practice requires a system that generalizes reliably across heterogeneous deployment environments without retraining, a requirement that existing approaches have yet to satisfy~\cite{gu2025csipose, xia2024self, chen2025towards, Liu2024PrivacySensing, Hinojosa2022PrivHAR}.

The fundamental barrier to such cross-environment robustness lies in the physics of wireless propagation. Wireless signals are highly sensitive to environmental conditions, including room layout, object placement, and transceiver configuration. Even small changes in these factors alter the multipath structure of the channel, causing severe domain shifts that rapidly degrade model performance in unseen scenarios~\cite{Zhou2023MetaFiPlus}. This sensitivity has three interconnected root causes. \textbf{First}, CSI measurements are dominated by reflections from static structures such as walls and furniture, whereas signal variations due to human motion are relatively weak. Pose-related information is therefore often obscured by strong, environment-dependent static components~\cite{Wang2022WiMesh}, making it difficult for a model trained in one environment to identify the same motion patterns in another. \textbf{Second}, the mapping from CSI to human pose is highly ambiguous. CSI signals represent superposed contributions from multiple propagation paths, so different body configurations can produce similar signal patterns under different multipath geometries. This many-to-many ambiguity means that a model lacking structural constraints will latch onto environment-specific correlations rather than learning truly transferable pose representations. \textbf{Third}, because the multipath geometry varies across environments, the distribution of both static and dynamic CSI components shifts substantially, creating a domain gap that purely data-driven adaptation strategies struggle to bridge without access to labeled target-environment data.

The common limitation of existing WiFi-based pose estimation methods is that they treat CSI as generic sequential or spatial data~\cite{Wang2019PersonInWiFi, Jiang2020WiPose, Zhou2022CSIFormer, Zhou2022PerUNet, Gian2024HPELi}, without explicitly accounting for the physical mechanisms that give rise to these three root causes. As a consequence, their learned representations inevitably mix static environmental components with dynamic human-induced signals, remain sensitive to multipath-induced ambiguity, and fail to generalize when the propagation structure changes. Addressing cross-environment robustness, therefore, requires a principled approach that models the underlying electromagnetic propagation rather than simply transplanting computer vision architectures to wireless signals.

To this end, we propose a physics-informed WiFi sensing framework for robust 3D human pose estimation under mobile and cross-environment settings. The key idea is to integrate physical insights from wireless signal propagation directly into the learning architecture, so that each component targets one of the three root causes identified above. Specifically, a physics-guided attention mechanism separates dynamic human-induced signal correlations from static environmental artifacts, thereby addressing the weak-motion signal problem. A hierarchical semantic fusion module introduces action-level structural constraints to resolve the ill-posed CSI-to-pose mapping. A physics-disentangled domain adaptation strategy explicitly decomposes the channel into static and dynamic components and enforces cross-environment invariance on the dynamic representation, enabling zero-shot generalization without target-environment labels. Together, these components form a unified framework grounded in electromagnetic propagation principles, enabling reliable pose estimation across diverse and previously unseen environments.
The main contributions of this work are summarized as follows:

\begin{itemize}

\item \textbf{Physics-Guided Multipath-Aware Attention (PGMA).} We propose a novel attention mechanism specifically designed for wireless sensing signals. Unlike standard self-attention, which treats inputs as generic sequences, the proposed PGMA explicitly incorporates antenna geometry, wavelength-dependent phase relationships, and multipath feasibility constraints into the attention formulation. This design enables the network to prioritize signal correlations consistent with electromagnetic propagation while suppressing environment-specific static artifacts, directly addressing the first root cause of cross-environment degradation.

\item \textbf{Hierarchical Cross-Attention Fusion (HCAF).} We introduce the first action-conditioned pose inference framework for WiFi sensing. By progressively integrating action-level semantic priors through dynamic convolution, cross-attention coupling, and joint-specific modulation, the proposed HCAF restricts pose estimation to biomechanically feasible submanifolds, thereby alleviating the many-to-many CSI-to-pose ambiguity that prevents learning of environment-invariant representations.

\item \textbf{Physics-Disentangled Domain Adaptation (PDDA).} We develop a principled domain adaptation strategy based on the physical decomposition of CSI signals. By separating static environmental components from dynamic human-induced perturbations and enforcing contrastive invariance on the dynamic representations, the proposed PDDA achieves robust cross-environment generalization without requiring labeled data in target environments.

\item \textbf{Theoretical Justification and Extensive Experiments.} We provide information-theoretic analysis showing that physics-guided representations increase mutual information with human pose, that action conditioning reduces conditional pose entropy, and that contrastive domain alignment promotes environment-invariant feature clustering. Extensive experiments on two public benchmarks and a real-world deployment dataset confirm that the proposed framework achieves state-of-the-art performance and strong zero-shot generalization across environments.

\end{itemize}

The remainder of this paper is structured as follows. Section~\ref{sec:related} reviews related work, and Section~\ref{sec:model} establishes the fundamental signal model and formally states the pose estimation problem. Subsequently, Section~\ref{sec:method} details the three core innovations of our proposed framework. To ground our design choices, Section~\ref{sec:theory} provides a theoretical analysis of the model's properties. Following this, Section~\ref{sec:experiments} reports comprehensive experimental results and concludes in Section~\ref{sec:conclusion}.

\section{Related Work}
\label{sec:related}

\subsection{Attention Mechanisms for WiFi HPE}

Early WiFi-based approaches employed convolutional neural networks to learn direct mappings from CSI to joint coordinates. Person-in-WiFi~\cite{Wang2019PersonInWiFi} pioneered end-to-end learning for this task, while WiPose~\cite{Jiang2020WiPose} extended the framework to 3D estimation through temporal modeling. However, the inherently local receptive fields of convolutional networks limit their ability to capture the global signal dependencies that characterize WiFi propagation. In practical environments, distant body movements can induce correlated variations across spatially separated antenna and subcarrier dimensions, which cannot be effectively modeled by local operators.

To address this limitation, recent work has explored attention mechanisms for modeling long-range dependencies. CSIFormer~\cite{Zhou2022CSIFormer} applies Transformer architectures to capture temporal correlations, while PerUNet~\cite{Zhou2022PerUNet} and HPE-Li~\cite{Gian2024HPELi} introduce channel-wise and kernel-based attention for spatial feature refinement. These approaches improve representation capacity and demonstrate the benefits of global modeling. However, they treat CSI as generic sequential or spatial data, without accounting for the physical structure of wireless propagation. In particular, antenna correlations depend on geometric configuration, subcarrier relationships are governed by wavelength-dependent patterns, and multipath propagation constrains feasible signal combinations. Because these methods operate on the full CSI observation without separating static environmental components from dynamic human-induced signals, the learned attention weights inevitably mix pose-irrelevant environmental artifacts with motion information, precisely the first root cause of cross-environment degradation. Although recent Transformer-based pose estimation methods leverage frequency-domain representations and structural priors~\cite{Li2023PoseFormerV2, Ci2023GFPOSE, Zhang2023MixSTE}, their direct application to WiFi signals remains suboptimal without modeling the underlying propagation mechanism. The proposed PGMA addresses this gap by introducing physics-guided attention that encodes antenna geometry and multipath characteristics into the attention formulation, enabling the network to concentrate on pose-relevant dynamic correlations.

\subsection{Semantic Guidance for Signal Ambiguity}

A fundamental challenge in WiFi-based human pose estimation lies in the ambiguous mapping between CSI measurements and human poses. Due to multipath propagation, different body configurations can produce similar signal patterns, while similar poses may correspond to distinct CSI observations under varying environments. Existing methods typically formulate this problem as an unconstrained regression task, directly predicting joint coordinates from signal features without incorporating high-level structural constraints~\cite{Wang2019PersonInWiFi, Jiang2020WiPose}. Without such constraints, models are susceptible to fitting environment-specific CSI patterns rather than learning transferable pose representations, which compounds the cross-environment generalization problem.

In contrast, the computer vision community has demonstrated the effectiveness of semantic and structural guidance for human pose estimation. Stacked hourglass networks~\cite{Newell2016Hourglass} and multi-resolution architectures such as HRNet~\cite{Sun2019HRNet} show that hierarchical feature fusion improves spatial reasoning. More recent work incorporates temporal consistency and structural priors to enhance robustness~\cite{Shan2024DiffPose, Zhang2023MixSTE}. Despite these advances, such semantic guidance has not been fully explored in WiFi sensing, where signal ambiguity is inherently more severe due to the indirect nature of observations and is further exacerbated by environmental variation. To address this gap, the proposed HCAF introduces a hierarchical fusion strategy that progressively injects action-level semantics into the feature representation, thereby constraining the solution space to physically plausible and biomechanically consistent poses and reducing environment-sensitive ambiguity in the CSI-to-pose mapping.

\subsection{Cross-Environment Domain Adaptation}

WiFi-based human pose estimation suffers from significant performance degradation across environments. Variations in room geometry, furniture layout, and electromagnetic interference alter the multipath propagation structure and lead to substantial domain shifts. As a result, models trained in one environment often fail to generalize to unseen settings, a central deployment challenge for mobile computing applications.

Recent work has explored domain adaptation techniques to mitigate this issue. MetaFi++~\cite{Zhou2023MetaFiPlus} and related meta-learning approaches~\cite{Radwan2025selfsupvised} improve adaptability by optimizing for rapid fine-tuning, but they require labeled samples from each target environment, which is impractical in real-world deployment. Adversarial domain adaptation methods based on gradient reversal~\cite{Ganin2016DANN, Tzeng2017ADDA} have shown effectiveness in vision tasks~\cite{Xu2023SelfSupervised}, yet their direct application to WiFi sensing is limited. In wireless systems, domain shift originates from changes in the physical signal generation process. Specifically, alterations in the multipath propagation structure rather than superficial differences in data distribution. Recent surveys~\cite{chen2023cross, ahmad2024wifi} confirm that WiFi domain variation is fundamentally tied to multipath propagation and environmental geometry. Purely data-driven adaptation strategies that ignore this physical origin therefore align the wrong signal components and provide only superficial invariance, leaving the root cause of cross-environment degradation unaddressed.

Existing work has made substantial progress in improving representation learning, structural modeling, and domain adaptation for WiFi-based human pose estimation. However, these advances have largely been pursued in isolation, and none provides a unified solution to all three root causes of cross-environment degradation: the dominance of static environmental components over weak motion signals, the many-to-many CSI-to-pose ambiguity, and the physics-rooted domain shift across environments. In contrast, this work proposes a unified physics-informed framework that simultaneously addresses all three root causes through propagation-aware attention, semantic guidance, and physically grounded domain adaptation, enabling reliable 3D pose estimation in practical mobile computing scenarios.

\section{System Model and Problem Formulation}
\label{sec:model}

\subsection{OFDM CSI Signal Model}

In OFDM-based WiFi systems, Channel State Information captures the complex frequency response between transmitter and receiver. For a single antenna pair at time $t$, the received signal follows $y = h(t) \cdot x + n$, where $x$ is the transmitted signal, $n$ represents additive noise, and $h(t)$ characterizes the time-varying channel response. The CSI vector aggregates the frequency response across $N_{\text{sub}}$ subcarriers as
\begin{equation}
\mathbf{h}(t) = \bigl[H(f_1,t),\, H(f_2,t),\, \ldots,\, H(f_{N_{\text{sub}}},t)\bigr]^T.
\end{equation}
Due to multipath propagation, the CSI at subcarrier $f_n$ is given by
\begin{equation}
H(f_n, t) = \sum_{k=1}^{M} a_k(t)\, e^{-j2\pi f_n \tau_k(t)},
\label{eq:csi}
\end{equation}
where $a_k(t)$ and $\tau_k(t)$ denote the complex gain and propagation delay of the $k$-th path, and $M$ is the total number of physical paths. Human motion modulates the channel by dynamically altering these components.

The key insight that drives the entire framework is that the multipath channel contains both time-invariant contributions from static scatterers and time-varying contributions induced by human motion. These two types of contributions convey fundamentally different information: the former encodes the environment geometry and is irrelevant to pose estimation, whereas the latter encodes the human body's kinematic state. To make this structure explicit, we decompose the channel response according to the temporal characteristics of individual multipath contributors as
\begin{align}
H(f_n,t) &= e^{-j2\pi\Delta f\cdot t} H_s(f_n) \notag \\
&\quad + \sum_{p=1}^{P} \rho_p(f_n,t)\, e^{-j\frac{2\pi d_p(t)}{\lambda_n}},
\label{eq:csi_decoup}
\end{align}
where $\Delta f$ is the carrier frequency offset between oscillators, $H_s(f_n)$ captures the time-invariant environmental contribution from static scatterers, $\rho_p(f_n,t)$ and $d_p(t)$ characterize the complex amplitude and instantaneous path length of the $p$-th dynamic component induced by human motion, and $\lambda_n = c/f_n$ denotes the wavelength at subcarrier $f_n$. For a system with $N_{\text{ant}}$ transmit-receive antenna pairs and $N_{\text{sub}}$ subcarriers, we construct the CSI matrix $\mathbf{H}_t \in \mathbb{C}^{N_{\text{ant}}\times N_{\text{sub}}}$ and extract amplitude-phase information into a real-valued tensor $\mathbf{X}_t \in \mathbb{R}^{L\times d_{\text{in}}}$, where $L = N_{\text{ant}} \times N_{\text{sub}}$ and $d_{\text{in}} = 2$.

The decomposition in~\eqref{eq:csi_decoup} reveals the key structure exploited by our framework. The static component $H_s(f_n)$ is estimated by temporal averaging over a window of $T_{\text{win}} = 10$ frames, which is long relative to human motion periods but short relative to environmental changes. The dynamic component $\mathbf{X}_d = \mathbf{X}_t - \mathbf{X}_s$ then concentrates the human-induced information that is relevant to pose estimation, while $\mathbf{X}_s$ encodes environment-specific propagation characteristics useful for domain calibration.

\begin{figure*}[t!]
\centering
\includegraphics[width=1.9\columnwidth]{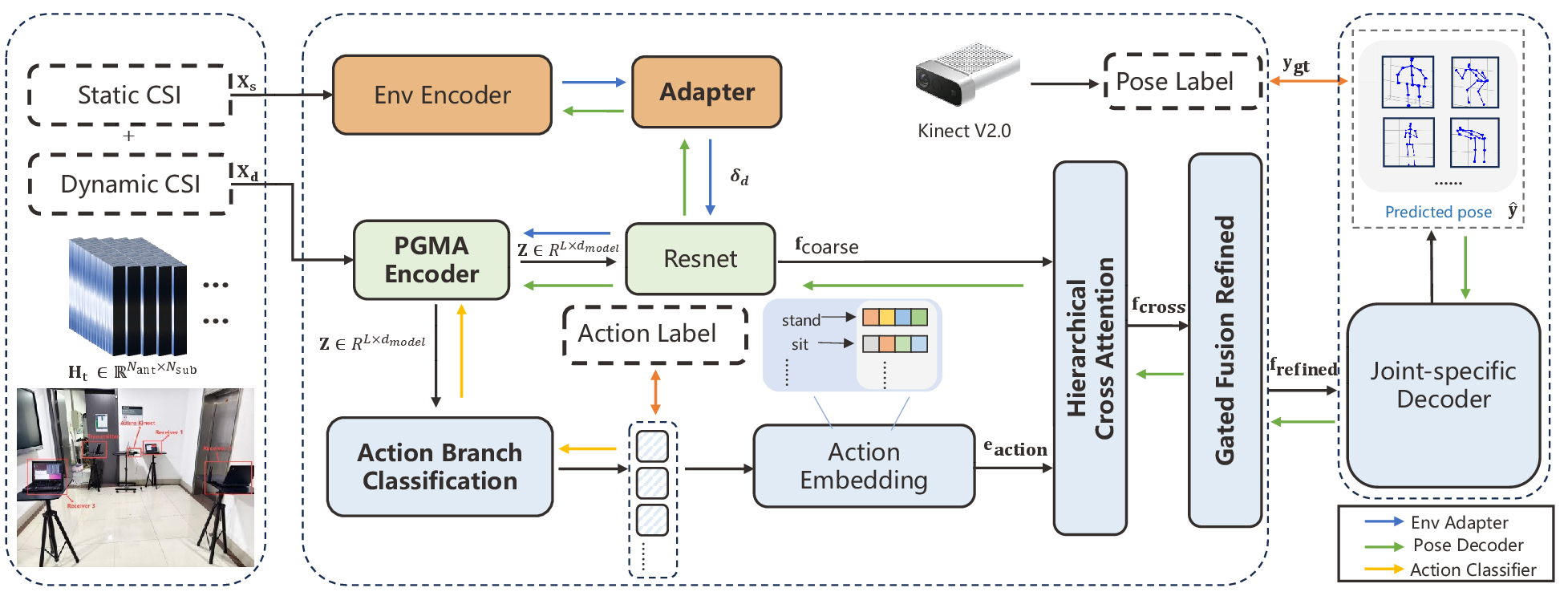}
\caption{Architecture of the proposed unified physics-informed framework for robust
WiFi-based 3D human pose estimation. The framework integrates three physics-grounded
components: Physics-Guided Multipath-Aware Attention for environment-invariant feature
extraction, Hierarchical Cross-Attention Fusion for action-guided ambiguity resolution,
and Physics-Disentangled Domain Adaptation for cross-environment generalization. The code and demonstration have been uploaded to GitHub at https://github.com/keysonhuang/Physical-Informed-WIFI-Sensing-for-3D-Human-Pose-Estimation. }
\label{fig:liuchentu}
\end{figure*}

\subsection{Problem Formulation}
\label{subsec:prob}
Building on the signal model above, we state the estimation problem in a form that exposes the three technical challenges that motivate the architecture. Let $\mathcal{D} = \{(\mathbf{X}_t^{(e)},\, a^{(e)},\, \mathbf{Y}^{(e)})\}_{e=1}^{E}$ denote the training set from $E$ environments, where $\mathbf{X}_t^{(e)}$ is the CSI observation, $a^{(e)} \in \{1,\ldots,N_{\text{action}}\}$ is the action label, and $\mathbf{Y}^{(e)} = \{\mathbf{y}_j\}_{j=1}^{J} \in \mathbb{R}^{J\times 3}$ is the 3D joint position ground truth with $J$ keypoints. We partition the model parameters into four functional groups $\boldsymbol{\theta} = (\boldsymbol{\theta}_{\text{enc}},\, \boldsymbol{\theta}_{\text{fus}},\, \boldsymbol{\theta}_{\text{dec}},\, \boldsymbol{\theta}_{\text{adp}})$, corresponding to the CSI encoder, the semantic fusion module, the pose decoder, and the environment adapters, respectively.

\begin{Prob}[Physics-Constrained WiFi-Based 3D HPE]
\label{prob:main}
Find parameters $\boldsymbol{\theta}^{\star}$ that solve
\begin{equation}
\min_{\boldsymbol{\theta}} \mathcal{L}_{\text{pose}} + \lambda_a\,\mathcal{L}_{\text{action}} + \lambda_d\,\mathcal{L}_{\text{domain}},
\label{eq:prob_obj}
\end{equation}
where the three loss terms are defined as
\begin{align}
\mathcal{L}_{\text{pose}} &= \frac{1}{|\mathcal{D}|} \sum_{e,t}\frac{1}{J}\sum_{j=1}^{J} \bigl\|\hat{\mathbf{y}}_j(\boldsymbol{\theta};\mathbf{X}_t^{(e)}) - \mathbf{y}_{\mathrm{gt},j}^{(e)}\bigr\|_2,
\label{eq:loss_pose}\\[3pt]
\mathcal{L}_{\text{action}} &= -\frac{1}{|\mathcal{D}|} \sum_{e,t}\sum_{c=1}^{N_{\text{action}}} \mathbf{1}[a^{(e)}=c]\log\hat{p}_c,
\label{eq:loss_action}\\[3pt]
\mathcal{L}_{\text{domain}} &= -\frac{1}{N_s}\sum_{i=1}^{N_s} \log\frac{\sum_{j\in\mathcal{P}(i)}\exp\!\bigl(\mathrm{sim}(\mathbf{z}_{d,i},\mathbf{z}_{d,j})/\tau\bigr)}{\sum_{k\neq i}\exp\!\bigl(\mathrm{sim}(\mathbf{z}_{d,i},\mathbf{z}_{d,k})/\tau\bigr)},
\label{eq:loss_domain}
\end{align}
with weights $\lambda_a = 0.5$ and $\lambda_d = 0.2$, contrastive temperature $\tau = 0.07$, $N_s$ the batch size, $\mathcal{P}(i)$ the set of samples sharing action $a^{(i)}$ but drawn from a different environment, $\mathrm{sim}(\cdot,\cdot)$ the cosine similarity, $\hat{p}_c$ the predicted probability for class $c$, and $\mathbf{z}_{d,i}$ the pooled dynamic feature of sample $i$.
\end{Prob}

The three objectives in~\eqref{eq:prob_obj} address distinct aspects of the estimation pipeline, and their interplay reflects the three challenges. The pose loss $\mathcal{L}_{\text{pose}}$ involves a highly non-convex, non-injective mapping from the high-dimensional CSI space to the 3D joint space, where identical CSI observations can arise from multiple distinct poses, and gradient descent without additional regularization converges to arbitrary local minima. The domain loss $\mathcal{L}_{\text{domain}}$ forces dynamic features from different environments to cluster toward a shared action-discriminative manifold, which must be balanced against the representational capacity required by $\mathcal{L}_{\text{pose}}$. The action loss $\mathcal{L}_{\text{action}}$ ensures that the semantic guidance branch maintains sufficient discriminability throughout training, as degraded action predictions propagate errors into the hierarchical fusion stages. The scalar weights $\lambda_a$ and $\lambda_d$ govern the Pareto trade-off among these competing objectives. 
The structure of Problem~\ref{prob:main} directly motivates the three-component architecture. The pose loss is minimized more effectively when the encoder produces representations concentrated on dynamic CSI components. The residual ambiguity in the CSI-to-pose mapping, which $\mathcal{L}_{\text{pose}}$ alone cannot eliminate, is addressed by $\mathcal{L}_{\text{action}}$ as additional information prior. The cross-environment generalization requirement captured by $\mathcal{L}_{\text{domain}}$ is handled by the domain adaptation. Each architectural decision thus corresponds to one term in the multi-objective formulation.

\section{Proposed Framework}
\label{sec:method}

\subsection{System Overview}
As shown in Fig.~\ref{fig:liuchentu}, the proposed framework addresses Problem~\ref{prob:main} through three physics-grounded components. Unlike existing methods that directly transplant computer vision architectures to wireless signals, each component is derived from the electromagnetic propagation model established in Section~\ref{sec:model}. The encoder first extracts discriminative features from the raw CSI tensor through physics-aware attention, the fusion module then resolves pose ambiguity by injecting action semantics at multiple granularities, and the domain adaptation layer ensures that the resulting representation generalizes across environments. We describe each component in turn.

\subsection{Physics-Guided Multipath-Aware Attention}
\label{subsec:pgma}

The physical decomposition in~\eqref{eq:csi_decoup} reveals that the full CSI observation $\mathbf{X}_t$ simultaneously contains static environmental components $\mathbf{X}_s$ and dynamic human-induced components $\mathbf{X}_d$. Since only $\mathbf{X}_d$ carries pose-relevant information while $\mathbf{X}_s$ constitutes environment-specific interference, an attention mechanism that concentrates on human-induced signal correlations while suppressing static artifacts is expected to yield substantially more discriminative representations. Standard self-attention operates on the full observation $\mathbf{X}_t$ without physical bias, attending equally to static and dynamic correlations alike. PGMA directly incorporates the physics prior through three coordinated mechanisms: electromagnetic positional encoding, learnable multipath feasibility masks, and frequency-decomposed band processing.


Each of the $L$ positions in $\mathbf{X}_t$ corresponds to one transmit-receive antenna pair and one subcarrier. For position $i$, let $d_i^{\text{ant}}$ denote the physical separation of the corresponding transmit-receive antenna pair and $f_i$ denote its subcarrier frequency, with wavelength $\lambda_i = c/f_i$. The electromagnetic positional encoding for position $i$ is
\begin{equation}
\mathbf{P}_i = \begin{bmatrix}
\sin(2\pi d_i^{\text{ant}}/\lambda_i)\\
\cos(2\pi d_i^{\text{ant}}/\lambda_i)\\
\sin(2\pi f_i/f_{\max})\\
\cos(2\pi f_i/f_{\max})
\end{bmatrix},
\label{eq:pe}
\end{equation}
where $f_{\max}$ is the maximum subcarrier frequency. The first two entries encode the spatial phase relationship governed by the antenna geometry, while the latter two capture the spectral progression across the OFDM band. Stacking encodings for all $L$ positions into $\mathbf{P}\in\mathbb{R}^{L\times 4}$, we inject them into input features as
\begin{equation}
\mathbf{X}_t^{\text{aug}} = \mathbf{X}_t + \mathbf{W}_P\,\mathbf{P},
\label{eq:aug}
\end{equation}
where $\mathbf{W}_P\in\mathbb{R}^{d_{\text{in}}\times 4}$ projects the encodings to the feature space. Following standard query-key-value projection, we modulate the attention scores with a learnable multipath feasibility mask:
\begin{equation}
A_{ij} = \frac{\exp\!\left(\dfrac{\mathbf{q}_i\mathbf{k}_j^T}{\sqrt{d_k}} + \log m_{ij}\right)}{\displaystyle\sum_{k=1}^{L} \exp\!\left(\dfrac{\mathbf{q}_i\mathbf{k}_k^T}{\sqrt{d_k}} + \log m_{ik}\right)},
\label{eq:pgma_attn}
\end{equation}
where $d_k$ is the key dimension and $m_{ij}\in(0,1]$ is the prior probability that positions $i$ and $j$ share significant multipath correlation. This formulation admits a Bayesian interpretation: the data-driven likelihood from the query-key product is combined with the physics-based log-prior $\log m_{ij}$ to approximate the posterior correlation probability under the multipath channel model. Using the per-position antenna separations $d_i^{\text{ant}}, d_j^{\text{ant}}$ and subcarrier frequencies $f_i, f_j$, the mask is initialized as
\begin{equation}
m_{ij}^{(0)} = \exp\!\left( -\frac{(d_i^{\text{ant}}-d_j^{\text{ant}})^2}{2\sigma_d^2} -\frac{(f_i-f_j)^2}{2\sigma_f^2}\right)
\label{eq:mask_init}
\end{equation}
and refined through training as $m_{ij} = \sigma(m_{ij}^{(0)}+\Delta m_{ij})$, where the learnable residual $\Delta m_{ij}$ allows adaptation to deployment-specific propagation characteristics, and $\sigma_d$ and $\sigma_f$ govern spatial and spectral coherence ranges respectively.

Human motion affects different frequency bands distinctly, as lower frequencies penetrate obstacles more effectively while higher frequencies provide finer spatial resolution. To exploit this spectral diversity, we partition the spectrum into $B=3$ frequency bands and process each with dedicated attention heads. The output of the $\ell$-th encoder layer is
\begin{equation}
\mathbf{Z}^{(\ell)} = \mathrm{LN}\!\left(\mathbf{Z}^{(\ell-1)} + \sum_{b=1}^{B} w_b^{(\ell)}\cdot \mathrm{PGMA}_b^{(\ell)}\!\left(\mathbf{Z}^{(\ell-1)}\right)\right),
\label{eq:band_fusion}
\end{equation}
where $\mathrm{LN}(\cdot)$ denotes layer normalization and the band weights
\begin{equation}
\mathbf{w}^{(\ell)} = \mathrm{Softmax}\!\left(\mathbf{W}_w^{(\ell)}\,\mathrm{GAP}\!\left(\mathbf{X}_t^{(\ell)}\right)\right)
\label{eq:band_weights}
\end{equation}
adapt to signal quality variations across bands, with $\mathrm{GAP}(\cdot)$ denoting global average pooling. Stacking $N_{\text{enc}}=4$ such layers with residual connections produces the refined feature representation $\mathbf{Z}\in\mathbb{R}^{L\times d_{\text{model}}}$ with $d_{\text{model}}=256$. This representation is subsequently passed to the HCAF module for action-guided ambiguity resolution.

\vspace{-1 pt}
\subsection{Hierarchical Cross-Attention Fusion}
\label{subsec:hcaf}

Even with well-extracted dynamic-component features, the CSI-to-pose mapping remains highly ambiguous because multiple distinct body configurations can produce similar CSI signatures through different multipath combinations. The action category to which a pose belongs constrains it to a specific biomechanical submanifold of the full pose space, thereby substantially narrowing the set of feasible configurations. HCAF operationalizes this constraint through three progressive stages of action semantic integration, moving from coarse action-conditioned modulation to fine-grained joint-specific adaptation.

First, the action semantics must be extracted to minimize the solution space. Global average pooling aggregates the PGMA features $\mathbf{Z}$ into a compact global descriptor $\mathbf{z}_{\text{global}} = \frac{1}{L}\sum_{i=1}^{L}\mathbf{Z}_i$. An action classifier predicts action probabilities
\begin{equation}
\hat{\mathbf{p}}_a = \mathrm{Softmax}\!\left(\mathbf{W}_2\,\mathrm{ReLU}(\mathbf{W}_1\,\mathbf{z}_{\text{global}} +\mathbf{b}_1)+\mathbf{b}_2\right),
\label{eq:action_pred}
\end{equation}
supervised by the cross-entropy loss $\mathcal{L}_{\text{action}}$ in~\eqref{eq:loss_action}. A continuous action embedding $\mathbf{e}_a = \mathbf{W}_{\text{emb}}\,\hat{\mathbf{p}}_a \in \mathbb{R}^{d_{\text{emb}}}$ serves as the semantic context for subsequent stages. This soft embedding preserves uncertainty in the action prediction, which is important for the robustness mechanism described at the end of this subsection.

Second, the hierarchical features are fused via action-conditioned modulation to strengthen the feature representation. The action embedding generates dynamic convolution kernels through a hypernetwork $\mathbf{k}_{\text{conv}} = \mathbf{W}_k\,\mathbf{e}_a$, reshaped into convolution weight format $\mathbf{K}_{\text{dyn}}$ and applied to the spatial rearrangement $\mathbf{Z}_{\text{sp}}$ of $\mathbf{Z}$:
\begin{equation}
\mathbf{f}_{\text{coarse}} = \mathrm{GAP}\!\left(\mathrm{ResNet}\!\left(\mathrm{Conv2D}(\mathbf{Z}_{\text{sp}},\,\mathbf{K}_{\text{dyn}})\right)\right).
\label{eq:coarse}
\end{equation}
Dynamic convolution kernels adapt the spatial feature extraction to the action context, allowing the network to emphasize anatomical regions that are most informative for the predicted action.

Lastly, refined Cross-Attention Coupling to obtain fine-grained features, thereby reducing ambiguity. The action embedding queries the coarse pose features via cross-attention:
\begin{equation}
\mathbf{f}_{\text{cross}} = \mathrm{Softmax}\!\left(\frac{\mathbf{Q}_a\,\mathbf{K}_{\text{pose}}^T}{\sqrt{d_k}}\right)\mathbf{V}_{\text{pose}},
\label{eq:cross_attn}
\end{equation}
where $\mathbf{Q}_a = \mathbf{W}_Q^c\,\mathbf{e}_a$, $\mathbf{K}_{\text{pose}} = \mathbf{W}_K^c\,\mathbf{f}_{\text{coarse}}$, and $\mathbf{V}_{\text{pose}} = \mathbf{W}_V^c\,\mathbf{f}_{\text{coarse}}$, with $d_k$ the key dimension shared with the PGMA encoder. A gating mechanism adaptively combines the cross-attended and coarse features:
\begin{align}
\mathbf{g} &= \sigma\!\left(\mathbf{W}_g\,[\mathbf{f}_{\text{coarse}},\,\mathbf{e}_a,\,\mathbf{f}_{\text{cross}}]\right),
\label{eq:gate}\\
\mathbf{f}_{\text{ref}} &= \mathbf{g}\odot\mathbf{f}_{\text{cross}} +(1-\mathbf{g})\odot\mathbf{f}_{\text{coarse}}.
\label{eq:refined}
\end{align}
The gate provides inherent robustness to action-prediction errors: when action predictions are uncertain, the gate increases reliance on the direct CSI-to-pose pathway, thereby reducing the weight of the semantic-guidance branch. Different body joints exhibit varying degrees of action dependence, so joint-specific adaptation layers allow each joint to selectively weight semantic information:
\begin{align}
\boldsymbol{\omega}_j &= \sigma(\mathbf{W}_j[\mathbf{f}_{\text{ref}},\,\mathbf{e}_a]),\\
\mathbf{f}_j &= \boldsymbol{\omega}_j\odot\mathbf{f}_{\text{ref}}.
\end{align}
Each joint is decoded independently through a two-layer MLP to produce the final 3D coordinates $\hat{\mathbf{y}}_j\in\mathbb{R}^3$. The joint-specific weights learned during training encode the anatomical prior that distal joints, such as wrists and ankles, are more strongly constrained by action context than proximal joints, such as the pelvis and spine.

\vspace{-3pt}
\subsection{Physics-Disentangled Domain Adaptation}
\label{subsec:pdda}

Environmental variations alter the static channel component $H_s(f_n)$ in~\eqref{eq:csi_decoup}, creating a domain shift rooted in propagation physics. This shift is fundamentally different from the appearance shift encountered in visual domain adaptation: whereas the latter can be addressed by aligning surface statistics, the former involves changes in the physical path structure of the channel that affect both the amplitude and phase relationships across all antenna-subcarrier pairs. Standard adversarial domain adaptation ignores this structure, confusing static and dynamic components that follow entirely different statistical behaviors. PDDA enforces environment-invariant dynamic representations by operating directly on the physically separated components.

We implement the decomposition in~\eqref{eq:csi_decoup} through temporal averaging over $T_{\text{win}}=10$ frames, yielding $\mathbf{X}_s$ and the residual $\mathbf{X}_d = \mathbf{X}_t - \mathbf{X}_s$. Dynamic features $\mathbf{z}_d = \mathrm{GAP}(\mathrm{PGMA}(\mathbf{X}_d;\boldsymbol{\theta}_{\text{enc}}))$ are subjected to the supervised contrastive loss $\mathcal{L}_{\text{domain}}$ in~\eqref{eq:loss_domain}: dynamic features from the same action cluster tightly across environments while features from different actions remain separable. The positive pairs in $\mathcal{P}(i)$ are deliberately drawn from different environments, so the contrastive loss directly rewards environment-invariant representations that preserve action discriminability.

Simultaneously, lightweight environment-specific adapters leverage the static component $\mathbf{X}_s$ to calibrate the decoder for deployment-specific propagation. For environment $e$, the adapter output $\boldsymbol{\delta}_e = \mathrm{Adapter}_e(\mathbf{X}_s)$, implemented as a two-layer MLP with 128 hidden units, is applied before decoding as
\begin{equation}
\mathbf{f}_{\text{adp}} = \mathbf{f}_{\text{ref}} + \boldsymbol{\delta}_e.
\label{eq:adapt}
\end{equation}
These adapters account for less than 1\% of total parameters, providing efficient environment-specific calibration without compromising the invariant dynamic representation. The PDDA design thus achieves a clean separation of concerns: the shared encoder and fusion modules learn environment-invariant pose features, while the adapters absorb the residual environment-specific signal that cannot be removed by physical decomposition alone.

\section{Theoretical Analysis}
\label{sec:theory}
In this section, we establish these properties as formal theorems, providing information-theoretic validation for the physics-guided attention and action-conditional fusion designs. We further analyze the contrastive domain loss to confirm that it promotes the environment-invariant feature geometry required by PDDA.

The theoretical analysis assumes that static environmental and dynamic human-induced components are approximately independent with respect to the pose variable. While strict independence may not hold in practical wireless environments, static multipath contributions typically evolve on a much slower timescale than human motion. Over short observation windows used for pose estimation, this temporal separation allows the dynamic components to capture the dominant pose-dependent signal variations. Consequently, the independence assumption provides a reasonable approximation for analyzing the representation properties of the proposed framework.

\subsection{Physics-Guided Representations and Mutual Information}

The decomposition in~\eqref{eq:csi_decoup} implies that $\mathbf{X}_t = \mathbf{X}_d + \mathbf{X}_s$, where $\mathbf{X}_s \perp \mathbf{Y}$ and $\mathbf{X}_s \perp \mathbf{X}_d$ hold by the independence of static scatterers from human motion. A generic representation $\mathbf{Z} = g(\mathbf{X}_t)$ will mix both components, potentially suppressing pose-relevant information. The following theorem quantifies the information cost of this mixing.

\begin{Thm}{\textbf{Physics-Guided Representation Maximizes Mutual Information:}}
\label{thm:mi}
Let $\mathbf{X}_t = \mathbf{X}_d + \mathbf{X}_s$ with $\mathbf{X}_s \perp \mathbf{Y}$ and $\mathbf{X}_s \perp \mathbf{X}_d$. For any measurable function $g$, let $\mathbf{Z}_d = g(\mathbf{X}_d)$ and $\mathbf{Z}_{\mathrm{mix}} = g(\mathbf{X}_d + \mathbf{X}_s)$. Then
\begin{equation}
I(\mathbf{Z}_d;\, \mathbf{Y}) \geq I(\mathbf{Z}_{\mathrm{mix}};\, \mathbf{Y}),
\label{eq:thm1}
\end{equation}
with equality if and only if $g(\mathbf{X}_d+\mathbf{X}_s) = \phi(g(\mathbf{X}_d))$ almost surely for some deterministic function $\phi$.
\end{Thm}

The proof is given in Appendix~A. The implication for the PGMA design is direct: the multipath feasibility mask $m_{ij}$ suppresses antenna-subcarrier pairs that cannot carry human-induced multipath, steering the learned representation toward $\mathbf{Z}_d$ and thereby increasing $I(\mathbf{Z};\mathbf{Y})$ relative to unconstrained attention. Standard attention applied to the full observation $\mathbf{X}_t$ does not satisfy the equality condition in Theorem~\ref{thm:mi}, because the static component $\mathbf{X}_s$ perturbs the query-key products in a way that depends on the environment rather than on $\mathbf{X}_d$ alone. It confirms a clear information-theoretic advantage of physics-guided design over generic attention.

It is instructive to translate this result into a quantitative bound. Let $\epsilon_s = I(\mathbf{X}_s;\mathbf{X}_d+\mathbf{X}_s|\mathbf{X}_d)$ measure the degree to which the static component perturbs the representation path. The chain rule applied and yields $I(\mathbf{Z}_d;\mathbf{Y}) - I(\mathbf{Z}_{\mathrm{mix}};\mathbf{Y}) \geq 0$, with the gap growing as the energy of $\mathbf{X}_s$ relative to $\mathbf{X}_d$ increases. In typical indoor environments, static multipath energy can exceed dynamic human-induced energy by an order of magnitude, implying a substantial practical benefit from physical decomposition prior to attention computation.

\subsection{Action-Conditional Inference and Entropy Reduction}

Theorem~\ref{thm:mi} addresses the feature extraction challenge but leaves the ambiguity problem: even with optimal dynamic-component features, the conditional entropy $H(\mathbf{Y}|\mathbf{X}_d)$ may remain high due to multipath superposition. The following theorem shows that this residual ambiguity is reduced by conditioning on the estimated action category, providing formal justification for the HCAF design in Section~\ref{subsec:hcaf}.

\begin{Thm}{\textbf{Action Conditioning Reduces Conditional Pose Entropy:}}
\label{thm:entropy}
Let $\hat{A}$ denote the action category estimated from $\mathbf{X}_t$. Then
\begin{equation}
H(\mathbf{Y}\mid\hat{A},\,\mathbf{X}_t) \leq H(\mathbf{Y}\mid\mathbf{X}_t),
\label{eq:thm2}
\end{equation}
with strict inequality whenever $I(\mathbf{Y};\hat{A}\mid\mathbf{X}_t)>0$. Furthermore, the entropy reduction is exactly
\begin{equation}
H(\mathbf{Y}\mid\mathbf{X}_t) - H(\mathbf{Y}\mid\hat{A},\mathbf{X}_t) = I(\mathbf{Y};\hat{A}\mid\mathbf{X}_t) \geq 0.
\label{eq:entropy_reduction}
\end{equation}
\end{Thm}

The proof is given in Appendix~B. The strict inequality holds whenever the action classifier retains residual information about $\mathbf{Y}$ beyond what is already captured by $\mathbf{X}_t$ alone. As established in the proof, this condition is satisfied precisely when the estimated action is informative about the biomechanical submanifold of the pose, which holds for any classifier with accuracy exceeding random chance on at least two action classes.

The identity in~\eqref{eq:entropy_reduction} further implies that the entropy reduction is monotone in classifier quality. Let $A$ denote the true action label. Since $\hat{A}$ is a stochastic function of $A$, the data processing inequality gives $I(\mathbf{Y};\hat{A}|\mathbf{X}_t) \leq I(\mathbf{Y};A|\mathbf{X}_t)$, with equality when the classifier is perfect. This monotonicity directly motivates the design of cross-entropy supervision in $\mathcal{L}_{\text{action}}$: maximizing classifier accuracy maximizes the entropy reduction achieved by conditioning, thereby reducing ambiguity in pose regression. Conversely, when classifier confidence is low, the gating mechanism in~\eqref{eq:gate} down-weights the action-conditioned pathway, reverting to direct pose regression. 

\subsection{Contrastive Domain Alignment and Feature Separability}

The previous two theorems address feature extraction and ambiguity resolution, but leave the question of whether the contrastive loss $\mathcal{L}_{\text{domain}}$ in~\eqref{eq:loss_domain} actually promotes environment-invariant representations. The following result establishes that minimizing $\mathcal{L}_{\text{domain}}$ is equivalent to maximizing a lower bound on the mutual information between dynamic features and action labels, while simultaneously minimizing the mutual information between dynamic features and the environment identity.

\begin{Thm}{\textbf{Contrastive Alignment Promotes Environment-Invariant Representations:}}
\label{thm:domain}
Let $E$ denote the environment index and $A$ the action label. For any dynamic feature representation $\mathbf{z}_d$, define the environment leakage as $I(\mathbf{z}_d; E)$ and the action informativeness as $I(\mathbf{z}_d; A)$. Minimizing $\mathcal{L}_{\text{domain}}$ over the encoder parameters $\boldsymbol{\theta}_{\text{enc}}$ simultaneously increases a lower bound on $I(\mathbf{z}_d; A)$ and decreases an upper bound on $I(\mathbf{z}_d; E \mid A)$, approaching the ideal representation satisfying $\mathbf{z}_d \perp E \mid A$ as the loss approaches its minimum.
\end{Thm}

To see why this holds, note that the contrastive loss in~\eqref{eq:loss_domain} constructs positive pairs by selecting samples that share the same action across different environments. A representation that minimizes this loss must assign high similarity to same-action pairs from different environments and low similarity to different-action pairs regardless of environment. The first condition directly drives $I(\mathbf{z}_d; E \mid A)$ toward zero, because features that distinguish environments within the same action class will contribute to misranking. The second condition drives $I(\mathbf{z}_d; A)$ upward, because action-discriminative features are required to separate different-action pairs in the contrastive objective. The combined effect therefore approaches the conditional independence $\mathbf{z}_d \perp E \mid A$, which is precisely the invariance property required for zero-shot generalization across environments. A formal proof following the InfoNCE lower-bound argument is provided in Appendix~C.


\section{Experimental Evaluation}
\label{sec:experiments}

\subsection{Experimental Setup}

\subsubsection{Datasets}

\begin{figure}[t!]
\centering
\includegraphics[width=0.8 \columnwidth]{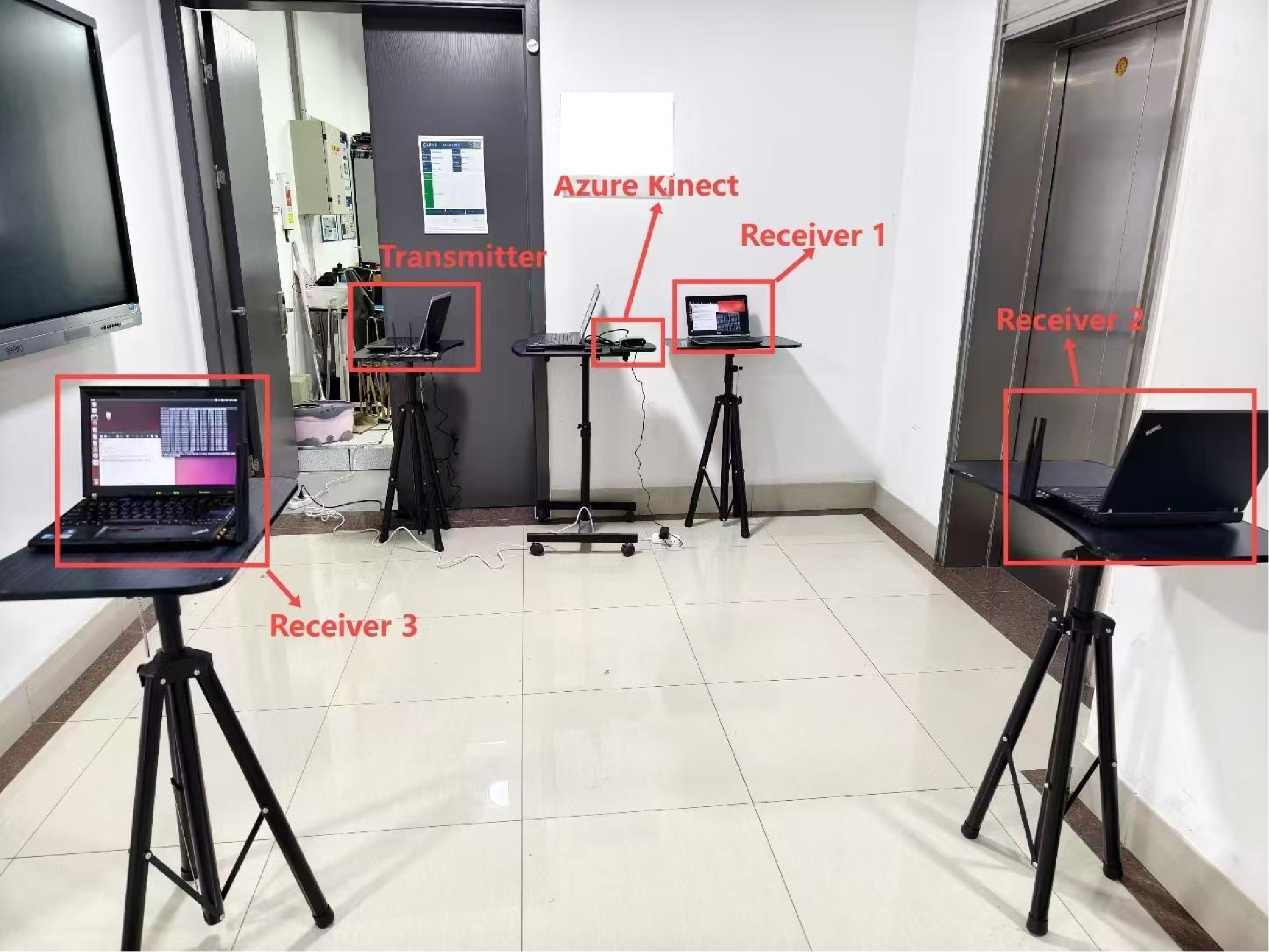}
\caption{Real-world data collection setup showing the WiFi transceiver configuration and a synchronized camera system. One transmitter and three distributed receivers with three antennas each create nine independent CSI links, each capturing 30 OFDM subcarriers at 5.32 GHz.}
\label{fig:data_collection}
\vspace{-10pt}
\end{figure}

We evaluate on two publicly available benchmarks and one self-collected dataset. \textbf{Person-in-WiFi-3D}~\cite{Yan2024PersonInWiFi3D} is the first end-to-end multi-person 3D pose estimation dataset using WiFi signals, containing data from seven volunteers performing eight daily actions across three locations with approximately 90,000 training and 7,800 test samples annotated with 14-keypoint 3D poses. \textbf{MM-Fi}~\cite{Yang2023MMFi} is a large-scale multi-modal benchmark featuring 40 subjects performing 27 actions in four environments across 320,000 frames with 17-keypoint annotations, it defines three evaluation protocols for daily activities (P1), rehabilitation exercises (P2), and all activities (P3), along with three split settings for random (S1), cross-subject (S2), and cross-environment (S3) evaluation. The \textbf{Real-World Prototype Dataset} was collected using an Intel 5300 NIC with one transmitting antenna and three distributed receivers, illustrated in Fig.~\ref{fig:data_collection}, each with three receiving antennas, providing nine independent CSI links across 30 OFDM subcarriers at 5.32 GHz. It covers eight actions across two indoor scenarios, yielding 28,406 aligned frames with CSI sampled at 600 Hz and ground truth captured at 30 fps.

\begin{table}[t!]
\centering
\caption{Results on the Person-in-WiFi-3D Dataset~\cite{Yan2024PersonInWiFi3D}.}
\label{tab:personinwifi}
\setlength{\tabcolsep}{3pt}
\footnotesize
\begin{tabular}{lcccc}
\toprule
\textbf{Method} & \textbf{PCK@20} & \textbf{PCK@50} & \textbf{MPJPE} & \textbf{PA-MPJPE} \\
\midrule
WiPose~\cite{Jiang2020WiPose}       & 61.5 & 87.8 & 118.9 & 70.5 \\
CSIFormer~\cite{Zhou2022CSIFormer}   & 63.0 & 87.9 & 115.1 & 71.9 \\
PerUNet~\cite{Zhou2022PerUNet}       & 63.5 & 88.5 & 113.9 & 71.1 \\
MDPose~\cite{Tang2023MDPose}         & 64.1 & 88.0 & 112.5 & 70.3 \\
MetaFi++~\cite{Zhou2023MetaFiPlus}  & 65.2 & 89.8 & 110.5 & 68.9 \\
PowerSkel~\cite{Yin2024PowerSkel}   & 66.1 & 90.1 & 108.7 & 67.9 \\
Person-in-WiFi-3D~\cite{Yan2024PersonInWiFi3D} & 71.5 & 91.5 & 91.7 & 57.2 \\
HPE-Li~\cite{Gian2024HPELi}         & 71.4 & 91.7 & 92.0  & 57.9 \\
DT-Pose~\cite{chen2025towards}      & 73.0 & 91.6 & 90.0  & 58.7 \\
\midrule
\textbf{Ours} & \textbf{74.1} & \textbf{92.0} & \textbf{84.98} & \textbf{55.5}\\
\bottomrule
\end{tabular}
\end{table}

\begin{table*}[ht!]
\centering
\caption{Comprehensive Results on the MM-Fi Dataset~\cite{Yang2023MMFi}.}
\label{tab:MMFiComprehensiveResults}
\resizebox{\textwidth}{!}{%
\begin{tabular}{l|cccc|cccc|cccc}
\hline
\multirow{2}{*}{\textbf{Method}} &
\multicolumn{4}{c|}{\textbf{P1}} &
\multicolumn{4}{c|}{\textbf{P2}} &
\multicolumn{4}{c}{\textbf{P3}} \\ \cline{2-13}
& PCK@20 & PCK@50 & MPJPE & PA-MPJPE
& PCK@20 & PCK@50 & MPJPE & PA-MPJPE
& PCK@20 & PCK@50 & MPJPE & PA-MPJPE \\ \hline\hline
\multicolumn{13}{l}{\textbf{S1 (Random Split)}} \\ \hline
MetaFi++~\cite{Zhou2023MetaFiPlus}
  & 48.7 & 86.2 & 187.2 & 120.5 & 32.0 & 81.2 & 214.3 & 121.6
  & 43.5 & 84.7 & 197.5 & 120.7 \\
HPE-Li~\cite{Gian2024HPELi}
  & 53.2 & 88.2 & 176.5 & 110.8 & 35.2 & 84.2 & 208.0 & 110.7
  & 48.1 & 86.9 & 185.4 & 107.2 \\
DT-Pose~\cite{chen2025towards}
  & \textbf{57.6} & 88.8 & 166.5 & 104.1
  & \textbf{40.1} & 82.5 & 196.7 & 102.6
  & 50.8 & 86.4 & 178.9 & 104.8 \\
Ours
  & 54.2 & \textbf{92.3} & \textbf{156.2} & \textbf{72.4}
  & 38.8 & \textbf{88.0} & \textbf{185.8} & \textbf{73.4}
  & \textbf{52.3} & \textbf{92.1} & \textbf{157.8} & \textbf{70.9} \\ \hline
\multicolumn{13}{l}{\textbf{S2 (Cross-Subject)}} \\ \hline
MetaFi++~\cite{Zhou2023MetaFiPlus}
  & 35.4 & 82.5 & 224.3 & 123.1 & 25.0 & 78.6 & 247.6 & 123.0
  & 33.3 & 81.1 & 232.2 & 124.2 \\
HPE-Li~\cite{Gian2024HPELi}
  & 39.1 & 85.8 & 220.1 & 105.8 & 27.2 & 78.2 & 243.3 & 102.5
  & 36.5 & 81.5 & 228.6 & 108.7 \\
DT-Pose~\cite{chen2025towards}
  & 41.1 & 85.9 & 215.1 & 106.6 & 28.1 & 78.5 & 240.3 & 100.2
  & 37.2 & 82.1 & 223.6 & 107.3 \\
Ours
  & \textbf{43.9} & \textbf{89.8} & \textbf{190.3} & \textbf{76.6}
  & \textbf{35.5} & \textbf{86.8} & \textbf{203.4} & \textbf{74.3}
  & \textbf{41.3} & \textbf{89.4} & \textbf{190.9} & \textbf{75.2} \\ \hline
\multicolumn{13}{l}{\textbf{S3 (Cross-Environment)}} \\ \hline
MetaFi++~\cite{Zhou2023MetaFiPlus}
  & 4.3 & 45.1 & 370.8 & 122.1 & 4.8 & 40.9 & 365.2 & 118.3
  & 4.5 & 40.1 & 372.5 & 118.2 \\
HPE-Li~\cite{Gian2024HPELi}
  & 5.6 & 47.2 & 361.1 & 110.3 & 5.3 & 42.3 & 358.2 & 104.0
  & 5.1 & 42.9 & 365.4 & 115.9 \\
DT-Pose~\cite{chen2025towards}
  & \textbf{10.2} & \textbf{58.5} & \textbf{334.5} & 106.2
  & 4.3 & 48.6 & 340.3 & 103.4
  & \textbf{8.9} & \textbf{60.4} & \textbf{320.8} & 106.2 \\
Ours
  & 7.2 & 57.6 & 335.7 & \textbf{78.9}
  & \textbf{6.1} & \textbf{51.3} & \textbf{331.3} & \textbf{77.8}
  & 5.7 & 54.3 & 335.5 & \textbf{80.4} \\ \hline
\end{tabular}%
}
\end{table*}

\subsubsection{Evaluation Metrics}

The Mean Per Joint Position Error (MPJPE) measures the average Euclidean distance $\mathrm{MPJPE} = \frac{1}{J}\sum_{j}\|\hat{\mathbf{y}}_j - \mathbf{y}_{\text{gt},j}\|_2$ in millimeters and serves as the primary accuracy metric. Procrustes Aligned MPJPE (PA-MPJPE) applies an alignment transformation removing global rotation, translation, and scale, evaluating structural pose quality independently of global positioning. The Percentage of Correct Keypoints
$\mathrm{PCK}@\alpha = \frac{1}{J}\sum_j\mathbb{I}\!\left(\frac{\|\hat{\mathbf{y}}_j - \mathbf{y}_{\text{gt},j}\|_2}{\|\mathbf{p}_{r_s}-\mathbf{p}_{l_h}\|_2} \leq \alpha\right)$
evaluates joint localization within a torso-normalized threshold, where $\mathbf{p}_{r_s}$ and $\mathbf{p}_{l_h}$ denote the right shoulder and left hip positions respectively.

\subsubsection{Implementation Details}

All experiments are implemented in PyTorch 2.0 on NVIDIA RTX A6000 GPUs. The PGMA encoder stacks $N_{\text{enc}}=4$ Transformer layers with $h=8$ attention heads and $d_{\text{model}}=256$, using feed-forward dimension $d_{\text{ff}}=1024$ and dropout 0.1. The action embedding dimension is $d_{\text{emb}}=128$. The Adam optimizer uses $\beta_1=0.9$, $\beta_2=0.999$, initial learning rate $1\times10^{-4}$ decaying by a factor of 0.5 every 30 epochs, batch size 32, and 100 training epochs. The total model has approximately 13.87M parameters and achieves an inference speed of 258.4 FPS, corresponding to 0.039 GFLOPs per frame.

\begin{table}[t!]
\centering
\caption{Per-joint MPJPE on the MM-Fi dataset.}
\label{tab:mmfi_joint}
\small
\setlength{\tabcolsep}{5pt}
\begin{tabular}{lrclr}
\toprule
Joint & MPJPE (mm) && Joint & MPJPE (mm) \\
\midrule
Pelvis  & 130.41 && Neck       & 158.15 \\
R-Hip   & 132.17 && Head       & 158.40 \\
R-Knee  & 133.17 && L-Shoulder & 152.25 \\
R-Ankle & 134.98 && L-Elbow    & 182.70 \\
L-Hip   & 131.40 && L-Wrist    & 239.00 \\
L-Knee  & 133.37 && R-Shoulder & 153.22 \\
L-Ankle & 138.43 && R-Elbow    & 182.00 \\
Spine   & 135.06 && R-Wrist    & 238.30 \\
Thorax  & 149.60 &&            &        \\
\bottomrule
\end{tabular}
\end{table}

\begin{table}[t!]
\centering
\caption{Per-joint MPJPE on the Person-in-WiFi-3D dataset.}
\label{tab:person_joint}
\small
\setlength{\tabcolsep}{5pt}
\begin{tabular}{lrclr}
\toprule
Joint      & MPJPE (mm) && Joint      & MPJPE (mm) \\
\midrule
Head       & 47.5  && Neck/Torso & 41.5  \\
L-Shoulder & 89.5  && R-Shoulder & 64.5  \\
L-Elbow    & 132.0 && R-Elbow    & 245.5 \\
L-Hand     & 130.5 && R-Hand     & 175.5 \\
L-Hip      & 54.5  && R-Hip      & 30.5  \\
L-Knee     & 55.0  && R-Knee     & 28.5  \\
L-Foot     & 47.5  && R-Foot     & 47.2  \\
\bottomrule
\end{tabular}
\end{table}

\subsection{Performance on Two Public Datasets}
\subsubsection{The overall performance}
Table~\ref{tab:personinwifi} presents an evaluation on Person-in-WiFi-3D. Our method achieves the lowest MPJPE of 84.98 mm and PA-MPJPE of 55.5 mm, representing substantial improvements over the strongest baseline DT-Pose, which achieves 90.0 mm MPJPE. The method also attains the highest PCK@50 of 92.0\% and PCK@20 of 74.1\%. The magnitude of improvement is particularly notable for PA-MPJPE, reflecting that the physics-constrained representation and action-guided fusion maintain superior structural pose integrity across diverse body configurations.

Table~\ref{tab:MMFiComprehensiveResults} presents the full benchmark on MM-Fi under all three protocols and split settings. Under S1, our model achieves 156.2 mm MPJPE and 72.4 mm PA-MPJPE on P1, compared to DT-Pose at 166.5 mm and 104.1 mm, respectively. The PA-MPJPE advantage is especially pronounced across all protocols, reflecting that while absolute localization improves significantly, structural pose quality improves even more dramatically, as most competing methods remain above 100 mm in PA-MPJPE.

Under S2, our model yields 190.3 mm MPJPE and 76.6 mm PA-MPJPE in P1, compared to DT-Pose at 215.1 mm and 106.6 mm. The consistent margin across all three protocols demonstrates strong subject-independent learning, which we attribute to the action-guided fusion, which removes subject-specific appearance biases from the CSI representation.

The S3 setting presents the most challenging scenario. While DT-Pose achieves competitive absolute MPJPE in some protocols (334.5 mm in P1), its PA-MPJPE of 106.2 mm reveals severe structural degradation. Our model achieves a PA-MPJPE of 78.9 mm in P1, with consistent results of 77.8 mm and 80.4 mm in P2 and P3, demonstrating that physics-disentangled adaptation learns environment-invariant representations of human pose shape. This structural robustness under cross-environment evaluation provides the strongest validation of the PDDA component, consistent with the invariance guarantee established in Theorem~\ref{thm:domain}.

\subsubsection{Per-Joint Error Analysis}
Per-joint error analysis on MM-Fi, detailed in Table~\ref{tab:mmfi_joint}, reveals that core skeletal structures such as the pelvis and hip joints achieve the lowest errors at approximately 130 mm, benefiting from their large effective radar cross-section. Distal joints, including the wrists, display the highest variance at approximately 239 mm, reflecting the challenge of tracking rapidly articulating limbs with minimal reflective surfaces. Knee joints maintain accuracy comparable to proximal structures at approximately 133 mm, successfully capturing periodic lower-body dynamics.

On the Person-in-WiFi-3D benchmark in Table~\ref{tab:person_joint}, the model achieves exceptional precision in lower-body tracking with knee and hip errors below 65 mm, attributed to stable ground-level multipath propagation. Upper-limb estimation presents greater challenges, with the right elbow at 245.5 mm primarily due to occlusion angles during arm movement sequences, while overall hand tracking at approximately 175 mm remains within acceptable bounds for gesture recognition. The per-joint error profiles on both datasets are consistent with the physical intuition that joints with larger backscatter cross-sections and more stable propagation paths benefit most from the physics-guided representation.

\subsection{Real-World Deployment Validation}

\subsubsection{Numerical Results}

Table~\ref{tab:actual_result} presents zero-shot generalization results, where all methods are evaluated without additional training on the target environment. Our approach achieves 144.8 mm MPJPE and 105.5 mm PA-MPJPE, representing 10.8\% and 7.9\% improvements over DT-Pose, which achieves 162.3 mm and 114.5 mm, respectively. PCK@50 reaches 93.3\%. The PA-MPJPE advantage substantially exceeds all baselines, demonstrating that the physics-informed representation learns transferable pose structure rather than memorizing environment-specific propagation patterns.

\subsubsection{Per-Joint Analysis on Real-World Data}

Table~\ref{tab:real_joint} reveals that lower extremities demonstrate remarkable robustness, with ankle and knee joints achieving errors between 98 and 120 mm despite significant environmental noise, validating the practical utility for gait analysis. Error magnitudes increase progressively toward the distal upper extremities, with wrists exhibiting the largest deviations of 210 to 250 mm due to inherently weak backscatter from small cross-sectional areas, compounded by multipath interference and rapid Doppler shifts. The consistent per-joint error profile across controlled and uncontrolled environments confirms that the model learns anatomically meaningful representations rather than environment-specific artifacts.

\begin{table}[t!]
\centering
\caption{Performance on the real-world deployment dataset. All methods are evaluated without additional training on the target environment.}
\label{tab:actual_result}
\footnotesize
\begin{tabular}{lcccc}
\toprule
Method & PCK@20 & PCK@50 & MPJPE & PA-MPJPE \\
\midrule
MetaFi++~\cite{Zhou2023MetaFiPlus} & 64.3 & 85.6 & 189.7 & 129.2 \\
HPE-Li~\cite{Gian2024HPELi}        & 70.5 & 89.9 & 173.5 & 119.7 \\
DT-Pose~\cite{chen2025towards}     & 71.8 & 90.1 & 162.3 & 114.5 \\
\textbf{Ours} & \textbf{74.6} & \textbf{93.3} & \textbf{144.8} & \textbf{105.5} \\
\bottomrule
\end{tabular}
\end{table}

\begin{table}[t!]
\centering
\caption{Per-joint MPJPE on the real-world deployment dataset.}
\label{tab:real_joint}
\small
\setlength{\tabcolsep}{5pt}
\begin{tabular}{lrclr}
\toprule
Joint      & MPJPE (mm) && Joint      & MPJPE (mm) \\
\midrule
Head       & 138.5 && Neck       & 132.0 \\
R-Shoulder & 129.5 && R-Elbow    & 173.6 \\
R-Wrist    & 210.0 && L-Shoulder & 128.5 \\
L-Elbow    & 190.1 && L-Wrist    & 250.0 \\
R-Hip      & 116.5 && L-Hip      & 114.0 \\
R-Knee     & 112.5 && R-Ankle    & 98.5  \\
L-Knee     & 118.0 && L-Ankle    & 115.5 \\
\bottomrule
\end{tabular}
\end{table}

\begin{figure}[t!]
\centering
\includegraphics[width=0.9 \columnwidth]{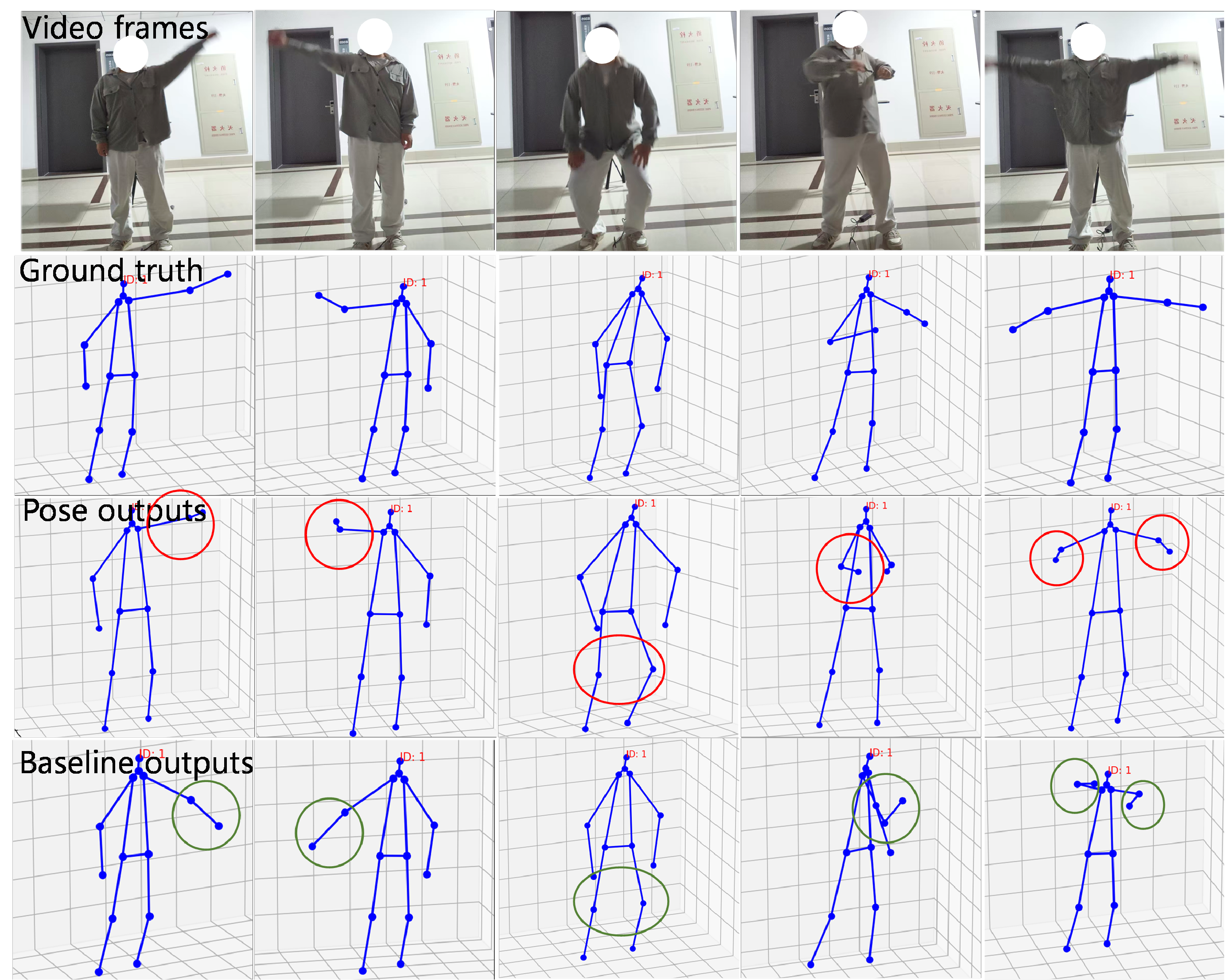}
\caption{Qualitative visualization on the real-world deployment dataset. The top row shows video frames, the second row shows ground truth, the third row shows our predictions with correct keypoints highlighted in red, and the bottom row shows baseline results with errors highlighted in green. Our method maintains anatomically plausible skeleton structures across diverse actions.}
\label{fig:output}
\vspace{-10pt}
\end{figure}

\subsection{Qualitative Visualization}

\begin{figure}[t!]
\centering
\includegraphics[width=0.9 \columnwidth]{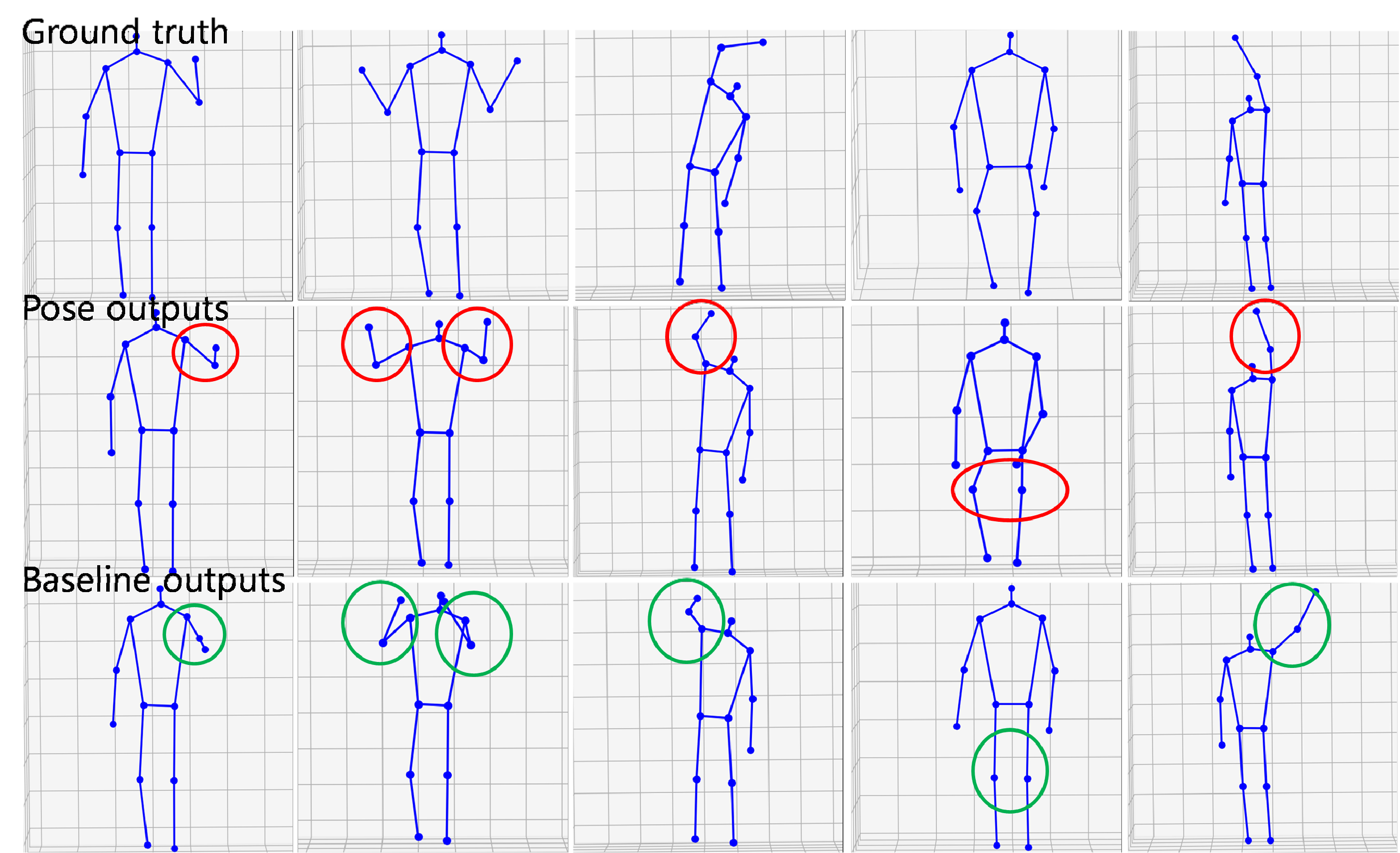}
\caption{Visualization on the Person-in-WiFi-3D dataset. The top row shows ground truth, the middle row our predictions (red circles), and the bottom row baseline results (green circles). Our method accurately tracks single-arm and bilateral gestures, while the baseline struggles to separate superposed arm signals.}
\label{fig:person_vis}
\vspace{-10pt}
\end{figure}

\begin{figure}[t!]
\centering
\includegraphics[width=0.9 \columnwidth]{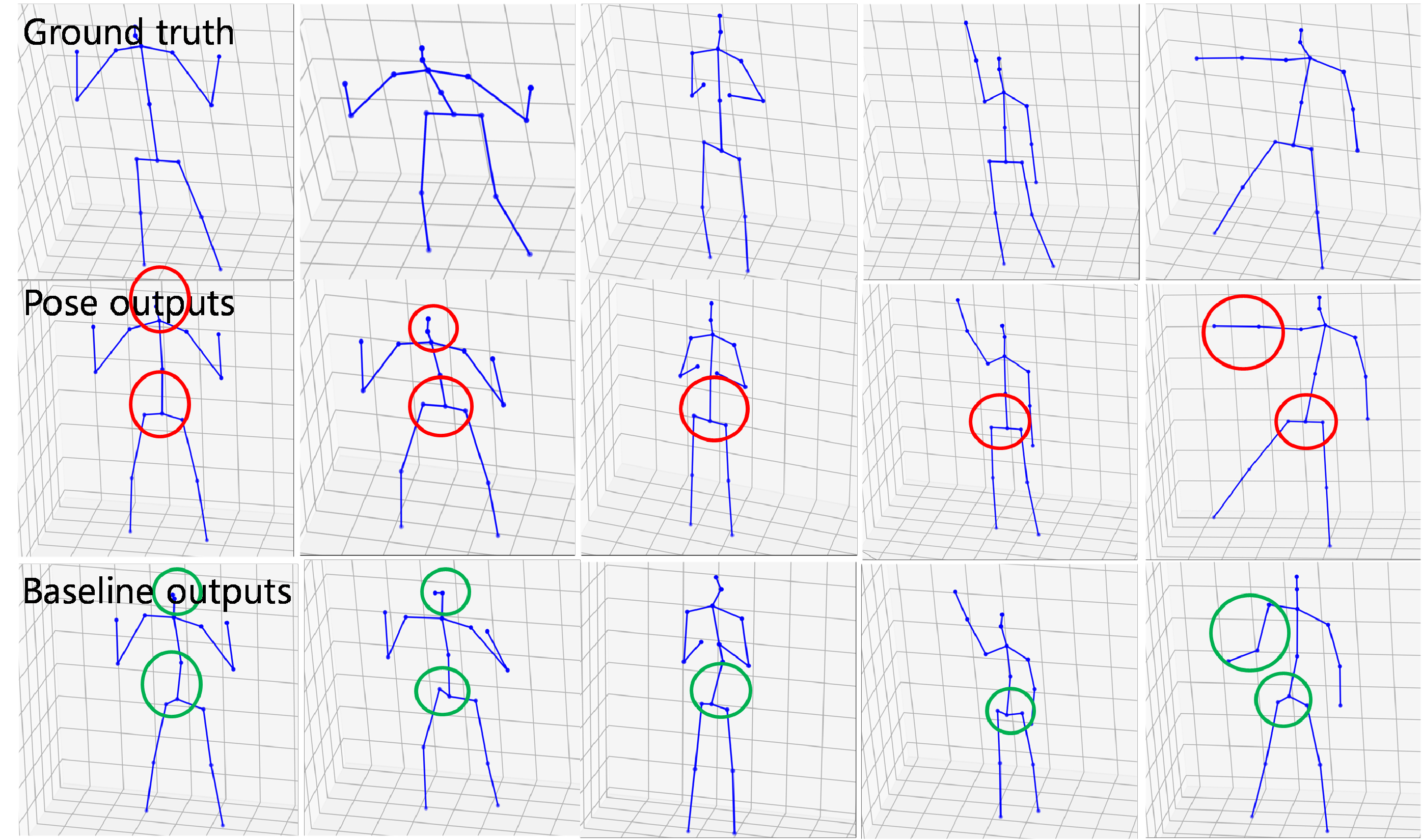}
\caption{Visualization on the MM-Fi dataset. The top row shows ground truth, the middle row our predictions (red circles), and the bottom row baseline results (green circles). Our approach captures full-body dynamics, including simultaneous limb abduction and asymmetric kicks, while the baseline produces collapsed or generic standing poses under high-amplitude motion.}
\label{fig:mmfi_vis}
\end{figure}

Fig.~\ref{fig:output} presents qualitative comparisons across the three evaluation datasets. On the real-world deployment data, our method generates pose estimations that closely align with the ground truth, maintaining anatomically plausible skeleton structures across all eight action categories. The baseline produces notable errors in joint localization, particularly in distorted limb positions and unrealistic joint configurations, as highlighted by green circles. On the Person-in-WiFi-3D data in Fig.~\ref{fig:person_vis}, our model correctly identifies unilateral and bilateral arm elevations, accurately capturing the V-shape configuration in the second column and the lateral extension in the third column, where the baseline underestimates the full range of motion. On the MM-Fi data shown in Fig.~\ref{fig:mmfi_vis}, which features highly dynamic actions, our approach successfully captures the simultaneous abduction of both arms and legs during jumping and maintains stable pelvic positioning during reaching movements, while the baseline exhibits collapsed poses with limbs appearing closer to the body.

\subsection{Ablation Studies}

\subsubsection{Component Contribution}

To validate the contribution of each architectural component, we conduct ablation experiments on MM-Fi protocols P3S1, P3S2, and P3S3 by systematically removing individual modules. As reported in Table~\ref{tab:ablation}, the full model consistently achieves the lowest MPJPE of 157.8 mm, 190.9 mm, and 335.5 mm across the three settings. The action branch contributes most significantly, as its removal causes 20.1\% degradation on P3S1 (from 157.8 mm to 189.5 mm), consistent with the entropy reduction predicted by Theorem~\ref{thm:entropy}. PGMA provides 7.0\% improvement by suppressing static environmental correlations, in agreement with Theorem~\ref{thm:mi}. Removing the learnable mask or the frequency band decomposition yields moderate increases of 4.3 mm and 5.0 mm on P3S1, respectively, confirming that both components contribute complementary signal-preprocessing benefits. The environment adapter yields a consistent 2.8\% improvement, validating the lightweight calibration strategy of PDDA and corroborating the invariance analysis of Theorem~\ref{thm:domain}.

\begin{table}[t!]
\centering
\caption{Ablation study results (MPJPE, mm) across three MM-Fi evaluation protocols, showing the individual contribution of each component. Lower is better.}
\label{tab:ablation}
\setlength{\tabcolsep}{13pt}
\footnotesize
\begin{tabular}{lccc}
\toprule
\textbf{Variant} & \textbf{P3S1} & \textbf{P3S2} & \textbf{P3S3} \\
\midrule
Ours (full model)            & \textbf{157.8} & \textbf{190.9} & \textbf{335.5} \\
w/o Learnable Mask           & 164.5          & 196.2          & 342.8          \\
w/o Freq. Decomposition      & 165.8          & 197.5          & 344.2          \\
w/o Env. Adapter             & 162.2          & 194.7          & 347.4          \\
w/o PGMA                     & 168.9          & 198.3          & 358.4          \\
w/o Action Branch            & 189.5          & 224.5          & 363.9          \\
\bottomrule
\end{tabular}
\end{table}

\subsubsection{Robustness to Imperfect Action Guidance}

Table~\ref{tab:action_percentage} evaluates performance under controlled action classification error injection. Performance degrades gracefully as the error rate increases, consistent with the gated fusion design: when action predictions are uncertain, the gate increases reliance on the direct CSI-to-pose pathway. Critically, even at 60\% action accuracy, the model achieves 186.8 mm MPJPE, still outperforming the 189.5 mm obtained without any semantic guidance. This confirms practical deployability under realistic action-recognition conditions and validates that the hierarchical fusion exploits action semantics when reliable, and falls back to direct regression when guidance quality is insufficient.

\begin{table}[t]
\centering
\caption{Robustness to imperfect action guidance (P3S1). Performance degrades gracefully, and even 60\% action accuracy surpasses the no-guidance baseline.}
\label{tab:action_percentage}
\setlength{\tabcolsep}{3pt}
\small
\begin{tabular}{ccccc}
\toprule
\textbf{Action Acc.} & \textbf{PCK@20} & \textbf{PCK@50} & \textbf{MPJPE} & \textbf{PA-MPJPE} \\
\midrule
100\%       & 52.3 & 92.1 & 157.8 & 70.9  \\
90\%        & 48.5 & 90.8 & 164.9 & 76.9  \\
80\%        & 46.8 & 89.4 & 172.4 & 83.2  \\
70\%        & 45.2 & 88.2 & 179.6 & 89.2  \\
60\%        & 43.6 & 86.8 & 186.8 & 95.2  \\
No guidance & 39.8 & 85.3 & 189.5 & 102.3 \\
\bottomrule
\end{tabular}
\end{table}

\subsection{Computational Complexity.}
The proposed framework is designed to maintain practical deployment efficiency despite incorporating physics-guided modules. We calculate the computational complexity for both the training and inference stages.

During training, the PGMA module introduces additional positional encodings and attention masks, which incur negligible overhead relative to standard self-attention operations. For sequence length $L = N_{\text{ant}} \times N_{\text{sub}}$ and model dimension $d_{\text{model}}$, computing query-key products requires $L^2 d_{\text{model}}$ multiply-accumulate operations per attention layer.
With typical configuration where $N_{\text{ant}} = 3$ and $N_{\text{sub}} = 30$ yielding $L = 90$ and $d_{\text{model}} = 256$, each attention layer requires approximately 2.07 million operations. With four encoder layers and eight attention heads, the total PGMA cost becomes approximately 66.3 million floating-point operations (mFLOPs).
The HCAF, including dynamic convolution generation, ResNet processing, cross-attention, and joint-specific decoding, requires approximately 350 mFLOPs. The PDDA adds minimal inference cost under 50 mFLOPs and employs environment-specific adapters comprising two linear layers, adding fewer than 3\% to the parameter count. 

During inference, the model has 13.87 million parameters and requires 0.039 GFLOPs per inference. On an NVIDIA RTX 6000 GPU, the system achieves approximately 258.4 frames per second for pose estimation, demonstrating real-time processing capability.

\section{Conclusion}
\label{sec:conclusion}

This paper presents a unified physics-informed deep learning framework for WiFi-based 3D human pose estimation that grounds each architectural decision in the underlying electromagnetic propagation model. The core problem is formulated as a multi-objective optimization problem that jointly minimizes the pose regression error, the action classification loss, and the contrastive domain alignment loss. The framework comprises three components: PGMA directly incorporates antenna geometry and multipath constraints into attention mechanisms, HCAF progressively integrates action-level semantic information to resolve the ill-posed CSI-to-pose mapping, and PDDA explicitly disentangles static environmental factors from dynamic human-induced signal variations to achieve cross-environmental robustness. Formal theoretical analysis confirms that physics-guided representations provably maximize mutual information with pose, that action conditioning provably reduces conditional pose entropy, and that the contrastive domain loss drives encoder features toward the ideal environment-invariant manifold. The framework achieves state-of-the-art performance on Person-in-WiFi-3D and MM-Fi and demonstrates strong zero-shot generalization on real-world deployment data.


\bibliographystyle{IEEEtran}
\bibliography{main}

\end{document}